\documentclass[aip, cha,reprint,amsmath,amsfonts]{revtex4-1}

\usepackage{amssymb}
\usepackage{amsmath}
\usepackage{bm}
\usepackage{paralist}
\usepackage{graphicx}
\usepackage{enumerate}
\usepackage{tikz}
\usepackage{graphicx}
\usepackage{verbatim}
\usepackage{etoolbox}
\usepackage{nicematrix}
\usepackage{booktabs}

\usetikzlibrary{matrix,positioning,fit}

\usepackage{hyperref}
\hypersetup{
	colorlinks,
	citecolor=blue,
	filecolor=blue,
	linkcolor=blue,
	urlcolor=blue,
	pdfproducer={}
}

\let\bbordermatrix\bordermatrix
\patchcmd{\bbordermatrix}{8.75}{4.75}{}{}

\newlength{\Mylen}

\usepackage{ragged2e}
\usepackage{caption}
\usepackage{ragged2e}
\DeclareCaptionJustification{justified}{\justifying}
\usepackage{soul}	

\usepackage[normalem]{ulem}

\usepackage[percent]{overpic}

\usepackage{xcolor}

\usepackage{cprotect}

\begin{document}

\title{Modeling density variations in two-dimensional microtubule-based active nematics}

\author{Kevin A.~Mitchell}
\email{kmitchell@ucmerced.edu}

\affiliation{Physics Department, University of California, Merced, CA
 95344, USA}

\author{Sean Ricarte}
\affiliation{Physics Department, University of California, Merced, CA
 95344, USA}

\author{Md Mainul Hasan Sabbir}
\affiliation{Physics Department, University of California, Merced, CA
 95344, USA}

\author{Brandon Klein}
 \affiliation{Department of Physics and Astronomy, Johns Hopkins University, Baltimore, MD 21218, USA.}

 \author{Daniel A. Beller}
 \affiliation{Department of Physics and Astronomy, Johns Hopkins University, Baltimore, MD 21218, USA.}

\date{\today}

\begin{abstract}
A dense two-dimensional layer of aligned microtubules (MTs), powered by molecular motors, is a canonical laboratory model of active materials  and a synthetic analog of biological systems such as bacterial turbulence, mitotic spindles, and morphogenesis.  This material exhibits nematic ordering and associated topological defects, which display complex emergent dynamics, including the creation and annihilation of defects and the braiding of defects around one another in a complicated chaotic dance.  Despite its prominent role in research, the MT-based active nematic material lacks a well established theoretical model that accurately captures the rich density variations prominently seen in experiments---density variations that are, in fact, the experimental signature of the nematic structure itself.  The MT-system is typically modeled using two fields: the Q-tensor (encoding the order and orientation of the
nematic phase) and the fluid velocity; critically, the microtubule
density is assumed to be constant.  This traditional model is adopted from classical Landau-de Gennes liquid crystal theory.  Here, we present a fundamentally different approach
to modeling MT-based active nematics that explicitly incorporates density variations,
producing simulations that strongly resemble experimental videos,
including the characteristic striation patterns.  It also reproduces important behavior of the system confined to a circular well---behavior seen experimentally, but not captured by current theory.  In crafting our model, we present an alternative to Landau-de Gennes theory for the creation and annihilation of topological defects that does not rely on the classic isotropic-nematic phase transition.
\end{abstract}

\maketitle

Active matter describes collective systems composed of many interacting subunits, each driven by a local energy source~\cite{Marchetti13,Das20}.  This energy drives the system into nonequilibrium dynamical states that exhibit complex emergent properties.  Examples of active matter can be either biological (swimming bacteria~\cite{Sokolov07,Wensink12,Dunkel13}, dynamic cell layers~\cite{Saw17,Kawaguchi17}) or synthetic (engineered biomolecular materials power by molecular motors~\cite{Ndlec97,Schaller10,Sanchez12,Henkin14,Giomi15,DeCamp15,Guillamat16,Shendruk17,Ellis18,Lemma19,Serra23,Tan19}) and can exist at microscopic (see prior examples) or macroscopic scales (flocks of birds~\cite{Toner95} and related collective animal motion~\cite{Katz11,Buhl06,Hueschen23}).  Active materials have drawn considerable research interest due to their complex material properties, nonlinear and nonequilibrium behavior, and capacity to mimic biological systems.  

\begin{figure} 
\includegraphics[width = 1\columnwidth]{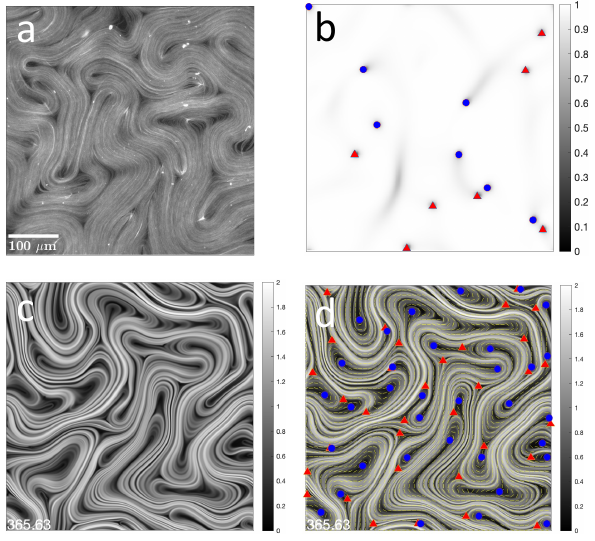}
\caption{\label{fig:Exp_Simulations} a) Fluorescence image of MT-based active nematic. Adapted from Ref.~\onlinecite{Tan19}, Fig.~2b.  Defects are marked with blue circles (+1/2) and red triangles (-1/2).  b) Simulation of active nematic using traditional Landau-de Gennes theory.  Shading represents the order parameter $S$.  Adapted from Ref.~\cite{Mitchell25}, Fig.~6a.  c) Simulations of nematohydrodynamic Eqs.~(\ref{r11}) and (\ref{r12}).  d) Same as c but showing director field.}
\end{figure}

One of the most studied and canonical active materials consists of a dense two-dimensional (2D) layer of microtubules (rod-like molecules) aligned parallel to each other in a nematic phase, exhibiting nearly perfect local orientational order~\cite{Sanchez12,Henkin14,DeCamp15,Serra23,Tan19} (Fig.~\ref{fig:Exp_Simulations}a).  This order breaks down at point-like topological defects, which come in two topological charges, $\pm 1/2$.  At $+1/2$ defects, the microtubules (MTs) bend around the defect in a hairpin turn.  At $-1/2$ defects, the MTs exhibit a three-fold structure.  The MTs are crosslinked by kinesin molecular motors that slide MTs relative to one another via the consumption of adenosine triphosphate (ATP).  This injection of energy causes the system to exhibit so-called ``active turbulence'' in which +1/2 and -1/2 topological defects are repeatedly created and destroyed in oppositely charged pairs, and  braid around one another in a complicated chaotic dance.  (The braiding perspective has been explored in Refs.~\onlinecite{Tan19,Smith22,Smith22b,Mitchell24,Firouznia25,Klein25}.)

All continuum models we are aware of for the canonical MT-based active nematic considered here assume a \emph{uniform} MT density and an incompressible flow.  (Some models of \emph{other} active nematic systems have included variable density~\cite{Ramaswamy03,Mishra14,Blow14,Gao15,Prost15,Julicher18,Nejad21,Pokawanvit22,Caballero22,Bhattacharyya23,Mirza25,Maddu26}.) 
However, the uniform-density assumption contradicts what one actually measures in experiments: pronounced density variations (Fig.~\ref{fig:Exp_Simulations}a.)  These variations are large, with the density plunging to 0 at topological defects.  Furthermore, the density variations exhibit prominent fractal striation patterns~\cite{Mitchell21}.  Finally, uniform density models have failed to model important experimental results, notably the behavior of the material when strongly confined to a circular well~\cite{Opathalage19,Hardouin22}.  The failure of current theory to reproduce such basic phenomena in a canonical laboratory system is a major shortcoming in the understanding of active materials.  

This letter addresses these deficiencies by presenting a continuum model of dense 2D active nematic MT-layers, producing striation patterns that closely match experiments.  Importantly, when confined to a circular well, this model reproduces both the doubly periodic state of Ref.~\onlinecite{Opathalage19} and the ``flagellar'' state of Ref.~\onlinecite{Hardouin22} via different  model parameters.  We construct this model from first principles and notably do not follow traditional Landau-de Gennes liquid crystal theory.  The broader significance is an alternative mechanism for the creation and annihilation of topological defects.

The standard approach to modeling MT-based active nematics is based on the theory of nematic liquid crystals~\cite{DeGennes93,Beris94}, which models not the density but the nematic order parameter $S$, along with the orientation of nematogens.   The order and orientation are packaged together in the Q-tensor.  The Q-tensor is coupled to the fluid flow and vice versa.  The Q-tensor is also governed by its free energy, which has the Landau-de Gennes form of an elastic energy plus a quartic phase energy  that favors the aligned nematic state ($S=1$) and disfavors the isotropic state ($S=0$).  Topological defects, at which $S = 0$, thus cost energy.  In this theory, defects are created when the local elastic energy is so large that it is energetically favorable to create a defect pair, where the material undergoes a nematic-to-isotropic phase transition at the defect.  Though the theory developed in this Letter shares some aspects of the liquid crystal theory, it deviates substantially by eliminating $S$ altogether.  Consequently, our theory avoids the phase free energy entirely.  Defects are created when the material fractures, allowing a void to open up in the density.

To create the striation pattern seen in experiments, two phenomena must occur~\cite{Mitchell21}.  First, density variations (e.g. voids and high density patches) must be formed at the scale of the defect spacing, and then these variations must be stretched and folded, pushing them down to smaller scales.  As evident from videos of the active nematic, during the stretching and folding process, there is essentially no diffusion, with individual bundles persisting over time, and material patches purely advected by the flow~\cite{Serra23}.  This fact is formalized in what we call the \emph{mass-locking principle}: the positions of the microtubule bundles are locked within the microtubule network and do not drift relative to the network, evolving only via coupling to the flow field of the network.  Of course, it is understood that individual microtubules stochastically attach and detach from the microtubule network.  The mass-locking principle assumes that the effect of such stochasticity on the large-scale structure of the network is slow relative to the dynamics of the flow.

We also adopt the \emph{nematic-locking principle}, introduced in Ref.~\onlinecite{Mitchell25}, which applies to the orientation of the microtubule bundles.  It states that the orientation of the director field evolves as a passive (infinitely thin) rod would in the fluid flow.  Equivalently, it implies that a nematic contour (i.e. an integral curve of the director field), passively advected in the flow, remains a nematic contour.
The nematic-locking principle holds throughout the majority of the material, but breaks down in regions where the material fractures (or heals), as when a defect pair is created (or destroyed).  As the microtubule density decreases, the bundles become weaker and eventually may fracture if bent sufficiently (Fig.~3d, Ref.~\onlinecite{Sanchez12}), allowing the director field to quickly relax.  The voids are formed in regions of high curvature, where the resulting large elastic pressure expands the material. 

\begin{figure} 
\includegraphics[width = 1\columnwidth]{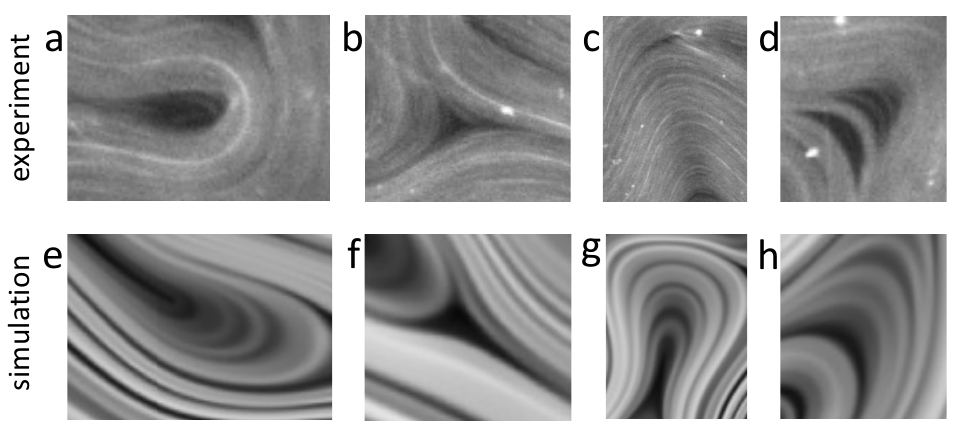}
\caption{\label{fig:CloseUps} Zoom of experimental images from Fig.~\ref{fig:Exp_Simulations}a: a) +1/2 defect, b) -1/2 defect, c) density suppression at bend, d) boomerang-shaped voids; e-h are corresponding features taken from the simulation in Fig.~\ref{fig:Exp_Simulations}c.  }
\end{figure}

We derive our theoretical model by first revisiting the definition of the Q-tensor.  Consider a two-dimensional microscopic ensemble of nematogens, i.e. rodlike particles, each with unit director $\mathbf{n}_i$.  Then construct the individual traceless q-tensor $\mathsf{q}_i = \mathbf{n}_i \otimes \mathbf{n}_i - \mathsf{I}/2$ for each.  The traditional Q-tensor, what we shall call $\mathsf{Q}_0$, is defined as the average of the $\mathsf{q}_i$'s within a small area $\delta A$, $\mathsf{Q}_0= \langle \mathsf{q} \rangle_N = \frac{1}{N} \sum_{i = 1}^N \mathsf{q}_i = S(\mathbf{n} \otimes \mathbf{n} - \mathsf{I}/2)$.  The last equality defines the order parameter $0 \le S \le 1$ and the macroscopic unit director field $\mathbf{n}$.  If the individual $\mathbf{n}_i$'s point in uniformly random directions, then $S = 0$, whereas if they all point along exactly the same line, $S = 1$.  Note that the normalization in the definition of $\mathsf{Q}_0$ is the number of nematogens $N$.  An alternative definition, what we shall call $\mathsf{Q}$ here, normalizes by the area, i.e.
\begin{equation}
    \mathsf{Q} = \langle \mathsf{q} \rangle_A = \frac{1}{\delta A} \sum_{i = 1}^N \mathsf{q}_i = \frac{N}{\delta A} \frac{1}{N} \sum_{i = 1}^N \mathsf{q}_i = \rho S(\mathbf{n} \otimes \mathbf{n} - \mathsf{I}/2),
\end{equation}
where $\rho$ is the number density.  There are two natural limits for $\mathsf{Q}$.  The first is the usual limit in which density goes to a constant value $\rho = 1$, in which case we recover $\mathsf{Q} = \mathsf{Q}_0$.  The second is the limit of perfect alignment in which $S$ goes to the constant value 1 and 
\begin{equation}
 \mathsf{Q} = \rho (\mathbf{n} \otimes \mathbf{n} - \mathsf{I}/2) = \rho \mathsf{P},  
 \label{r1}
\end{equation}
introducing the orientation tensor $\mathsf{P} = \mathbf{n} \otimes \mathbf{n} - \mathsf{I}/2$.  The limit $S = 1$ is much more appropriate for MT-based active nematics than $\rho = 1$, as demonstrated in Fig.~\ref{fig:CloseUps}a and b---near a defect core, the alignment remains nearly perfect ($S \approx 1$) while the density $\rho$ goes to 0, justifying Eq.~(\ref{r1}).


In addition to the Q-tensor, we need the velocity field $\mathbf{u}$.  We use a one-fluid model where $\mathbf{u}$ is the velocity of the dynamic network of crosslinked MTs.  The entrained aqueous background is not explicitly modeled, though it contributes to the effective viscosity of the flow of the MT network.  
The Q-tensor evolves via coupling to the flow $\mathbf{u}$, which moves and distorts the network of MTs, and via the dissipation of free energy, which fractures and realigns the network of MTs.  Note that we allow compressibility of the material.

As is well established (see Supplemental for a derivation), $\mathsf{P}$ evolves via coupling to the flow as
\begin{equation}
  \frac{D}{D t}  \mathsf{P}
 = \mathsf{E}' +  [\mathsf{P}, \Omega ] -
2 \text{Tr} (\mathsf{P} \mathsf{E}' ) \mathsf{P},  
\label{r7}
\end{equation}
where $D/Dt = \partial/ \partial t + \mathbf{u} \cdot \nabla$ is the advective derivative, $\Omega = (\nabla \mathbf{u} - \nabla \mathbf{u}^T)/2$ is the
vorticity tensor, and $\mathsf{E}'$ is the deviatoric (traceless) part of the rate-of-strain tensor $\mathsf{E} = (\nabla \mathbf{u} + \nabla \mathbf{u}^T)/2$.  Similarly, the density of a parcel in the Lagrangian frame changes as 
\begin{equation}
    \frac{D \rho}{Dt} = -\rho \nabla \cdot \mathbf{u}.
    \label{r22}
\end{equation}  

In addition to flow coupling, $\mathsf{P}$ evolves via fracturing and  realignment of the MTs, which we describe via gradient descent of a free energy $F = \int f(\mathsf{P},\nabla \mathsf{P},\rho,\nabla \rho) \; da$, 
\begin{equation}
    \frac{\partial P_{ij}}{\partial t} = - M_{ij, k \ell} \frac{\delta F}{\delta P_{k\ell}}, 
    \label{r8}
\end{equation}
using Einstein summation and where $M_{ij, k \ell}$ is the (symmetric, i.e. $M_{ij, k \ell} = M_{k \ell, ij}$) orientational mobility.
Since $|\mathbf{n}| = 1$,  $\partial \mathsf{P}/\partial t$ is proportional to $\mathsf{R} = \mathbf{n}^\perp \otimes \mathbf{n} + \mathbf{n} \otimes \mathbf{n}^\perp = \mathsf{J} \mathsf{P}$, where $\mathbf{n}^\perp = \mathsf{J} \mathbf{n}$ with $\mathsf{J}$ the right-handed rotation by $\pi/2$.  This in turn implies that the symmetric mobility tensor $M_{ij, k \ell}$ has the form $M R_{ij} R_{kl}$, with mobility scalar $M \ge 0$.  Thus, Eq.~(\ref{r8}) becomes
\begin{equation}
    \frac{\partial \mathsf{P}}{\partial t}  = M \mathsf{R} \text{Tr} (\mathsf{R h}),
\; \text{where} \;
\mathsf{h} = - \left(\frac{\delta F}{\delta \mathsf{P}} -\frac{\mathsf{I}}{2}\text{Tr}\frac{\delta F}{\delta \mathsf{P}}\right).
    \label{r5}
\end{equation}
Since the above term violates nematic-locking, which holds almost everywhere, the mobility $M$ is 0 almost everywhere.  It is only nonzero in regions of low density and/or large curvature, where the MTs can fracture and reorient easily, thereby relaxing the elastic energy.  We thus introduce a sigmoidal (i.e. a soft step function) switch function $f_{sw}(\rho, \kappa^2)$, which turns on the mobility in regions of low density and high (square) curvature $\kappa^2$.  Specifically, 
\begin{equation}
    f_{sw}(\rho,\kappa^2) = 
    \left[1 + e^{(\kappa^2_c(\rho) - \kappa^2)/ \kappa^2_w} \right]^{-1},
    \label{r21}
\end{equation}
with step sharpness $\kappa^2_w = 0.0025$, and where the critical value of $\kappa_c^2(\rho)$ at which fracturing occurs is also a sigmoid
\begin{equation}
    \kappa^2_c(\rho) = \kappa^2_\ell + (\kappa^2_u - \kappa^2_\ell)
    \left[1 + e^{(\rho_c - \rho)/ \rho_w} \right]^{-1},
\end{equation}
ranging from $\kappa^2_\ell = 0.025$ to $\kappa^2_u = 0.4$, with the step at $\rho_c = 0.3$ with sharpness  $\rho_w = \rho_c/10$.   Expressing the mobility as $M = \Gamma f_{sw}(\rho,\kappa^2)$, with \emph{rotational viscosity} $\Gamma$, Eqs.~(\ref{r7}) and (\ref{r5}) 
yield the full transport equation for $\mathsf{P}$,
\begin{equation}
  \frac{D}{D t}  \mathsf{P}
 = \mathsf{E}' +  [\mathsf{P}, \Omega ] -
2 \text{Tr} (\mathsf{P} \mathsf{E}' ) \mathsf{P} + \Gamma f_{sw}(\rho,\kappa^2)\mathsf{R} \text{Tr} (\mathsf{R h}).
\label{r11}
\end{equation}

In principle, we could also construct a mass current from $F$, but this would violate the mass-locking principle. Thus density evolves entirely via Eq.~(\ref{r22}).

The velocity $\mathbf{u}$ is determined by the Stokes equation
\begin{equation}
  \nabla_j \Pi_{ij} = 0,
\label{r12}
\end{equation}
since the Reynolds number is nearly 0, with stress tensor
\begin{equation}
  \Pi = 2\eta \mathsf{E}' + \xi \nabla \cdot \mathbf{u} +\Pi^E + \Pi^A,
  \label{r20}
\end{equation}
with shear viscosity $\eta$ and bulk viscosity $\xi$.  The elastic stress is (see Supplemental for a derivation) 
\begin{align}
  \Pi^E & = -\mathsf{h} +
  [\mathsf{P},\mathsf{h}] + 2 \text{Tr}(\mathsf{Ph}) \mathsf{P} \nonumber \\ &-\nabla P_{ij} \otimes \frac{\partial f}{\partial \nabla P_{ij}} -
  \nabla \rho \otimes \frac{\partial f}{\partial \nabla \rho} 
  +\left(f + \rho g \right) \mathsf{I},
  \label{r18}
\end{align}
where $g = -\delta F/ \delta \rho$.  Note that $\Pi^E$ includes both the deviatoric part ${\Pi^E}'$ and the isotropic, elastic pressure $-p \mathsf{I}$, where $p = -\text{Tr} (\Pi^E)/2$.  The fourth term in Eq.~(\ref{r20}) is the 
active stress driving the motion,
$
\Pi^A = -\zeta \rho^a \mathbf{n} \otimes \mathbf{n} = -\zeta \rho^a (\mathsf{P} + \mathsf{I}/2),
$
where $\zeta > 0$ is the (extensile) activity and the exponent $a$ is an empirical material property. 

We choose a free energy with elastic contribution only
\begin{equation}
    F = \int  \frac{K}{2} \rho^b \nabla_i P_{jk} \; \nabla_i P_{jk} dr.
    \label{r10}
\end{equation}
Note the absence of the quartic potential in $\mathsf{Q}$ central to Landau-de Gennes theory.  The exponent $b$ is an empirical material property.  Equation~(\ref{r10}) then yields
$\mathsf{h} = K \nabla_k (\rho^b \nabla_k \mathsf{P})$,  
$\frac{\partial f}{\partial \nabla \mathsf{P}} = K \rho^b \nabla \mathsf{P}$, 
$g = \frac{bf}{\rho}$,  $\frac{\partial f}{\partial \nabla \rho} = 0$.

Since elasticity scales like $K \rho^b$ and activity scales like $\zeta \rho^a$, the active length scales with density as $\rho^{(b-a)/2}\sqrt{K/\zeta}$.  Thus, if $a>b$ (we shall use $a = 2.4$ and $b = 2.0$), the effect of activity grows faster than elasticity as density increases.  Thus, higher density makes the material more modulationally unstable.  

We integrate Eqs.~(\ref{r22}), (\ref{r11}), and (\ref{r12}) on an $800 \times 800$ grid, with $\Delta x = 0.25$ and doubly periodic boundary conditions, beginning with a (nearly) horizontal director field, uniform initial density $\rho = 1$, and uniform velocity $\mathbf{u} = 0$.  We use parameter values $K = 90,112$, $\zeta = 4096$, $\eta = 2560$, $\xi = 75,000$, and $\Gamma = 1$.   Supplemental video S1 shows the simulation, with   Figure~\ref{fig:Exp_Simulations}c showing the final frame.  The simulation exhibits the same key features seen in prior simulations (and experiments)---defect creation and annihilation, chaotic defect motion, and active turbulence.  However, and this is the key result, we also generate density variations with voids that are stretched and folded into striation patterns, as in experiments.  Furthermore, the director field (Fig.~\ref{fig:Exp_Simulations}d) strongly aligns with the striation pattern. 

The simulations reproduce other experimental features.  First, the density profiles near +1/2 and -1/2 defects have the same characteristic teardrop and triangular shapes in both experiment and simulation (Fig.~\ref{fig:CloseUps}a,b,e,f).  Second, the density is lower at bends in the director field (Fig.~\ref{fig:CloseUps}c and g).  Third, both exhibit boomerang-shaped voids in curved regions (Fig.~\ref{fig:CloseUps}d and h).

\begin{figure}
    \centering
    \includegraphics[width=\linewidth]{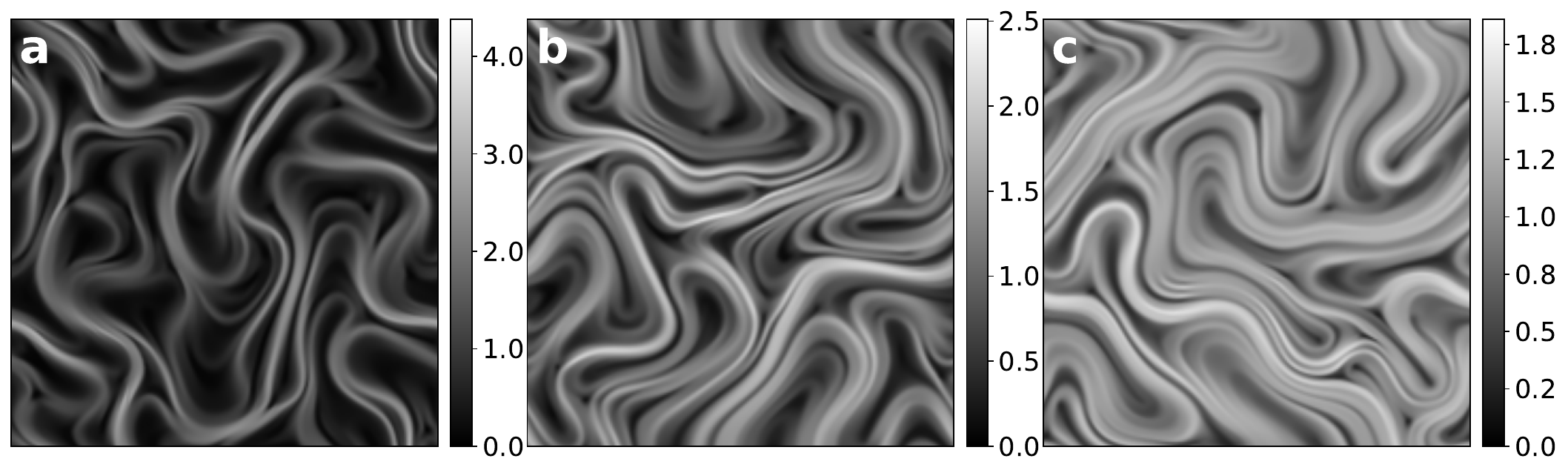}
    \caption{Simulation images on a $200 \times 200$ grid ($\Delta x = 1$)  with increasing bulk viscosity $\xi$. (a) $\xi = 2,560$, (b) $\xi = 25,600$, and (c) $\xi = 75,000$.  See Supplemental for more detail.}
    \label{fig:bulk-visc-comparison}
\end{figure}

Significant changes to the density pattern occur when the bulk viscosity is changed.  As the bulk viscosity is lowered, it is easier to compress the material and the resulting voids are larger (Fig.~\ref{fig:bulk-visc-comparison}).

As an application of the present model, we confine the material to a circular well using a force density $-\rho \nabla V_{\text{Well}}$, with 
$
    V_{\mathrm{well}}(r) = V_{\mathrm{0}}/ \left[ 1 + e^{-(r-R_{\mathrm{well}})/W_{\text{well}}} \right],
$
where $R_{\text{well}} = 60$ is the well radius, $W_{\text{well}} = 2$ is the well sharpness, and $V_0 = 10^5$ is the well height.  
Note that we omit any anchoring energy at the boundary, relying only on the well geometry for alignment. 
We discuss two sets of material parameters giving rise to distinct behavior seen in two distinct sets of experiments.

\begin{figure}
 \centering
  \includegraphics[width= 1 \columnwidth]{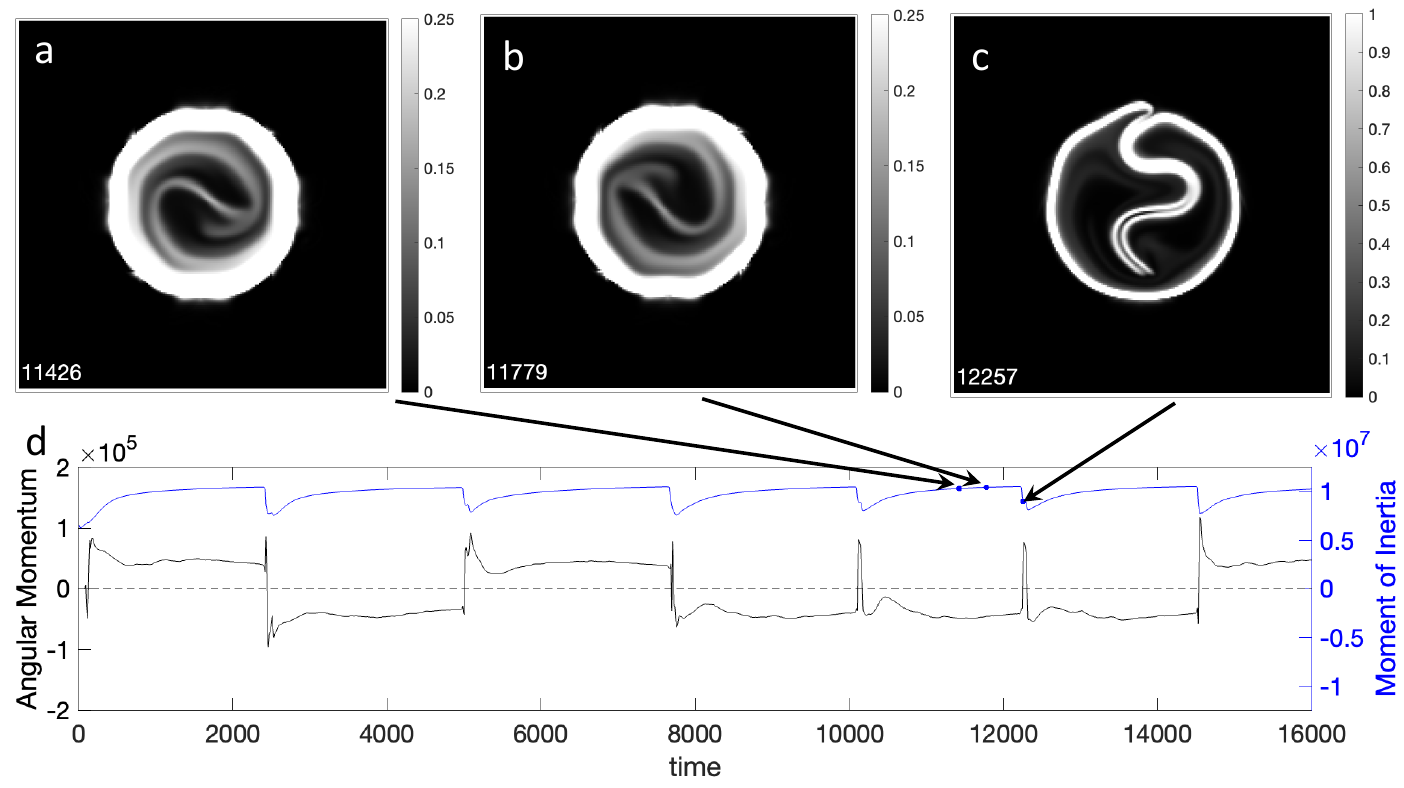}
  \caption{ Simulation of the doubly periodic state experimentally observed in Ref.~\onlinecite{Opathalage19}. (a) A two +1/2 defect state.  (b) The two defects have traded places, rotating a half-period about each other. (c)  A buckling event has injected mass into the interior. (d) Plots of the moment of inertia and angular momentum of the mass distribution, showing the bursting period and the changing sign of the angular momentum.}
  \label{fig:dblPer}
\end{figure}

First, we successfully observe the doubly periodic behavior seen experimentally in Ref.~\onlinecite{Opathalage19} using a $200 \times 200$ grid and the same material parameters as Fig.~\ref{fig:Exp_Simulations}c except a smaller total mass of $4,709$.  (See video S3.) Since $a>b$, the material becomes more modulationally unstable as density increases.  
Beginning with a uniform density inside the well and azimuthal director field, the system settles into the doubly periodic state, after an initial transient.  This behavior consists of two +1/2 defects that rotate around each other (Fig.~\ref{fig:dblPer}a and b), pushing material toward the wall.  As this progresses, the moment of inertia of the mass distribution rises  as a large void opens in the center.  Eventually, the annulus of material at the wall grows to sufficient mass that it becomes unstable and buckles, injecting a burst of material into the center of the well (Fig.~\ref{fig:dblPer}c).  This injected material again settles into the two defect state and the  process repeats indefinitely.  Thus there are two periods, the period of rotation of the two central defects and the bursting period at which  material is injected into the well.  Note that the sign of the angular momentum of the two-defect state changes in an apparently random fashion, similar to the experiments~\cite{Opathalage19}.  One difference from the experiments presented in Ref.~\onlinecite{Opathalage19} is that the simulated mass injection events are more chaotic and contain more defects.

\begin{figure}
 \centering
  \includegraphics[width= 1 \columnwidth]{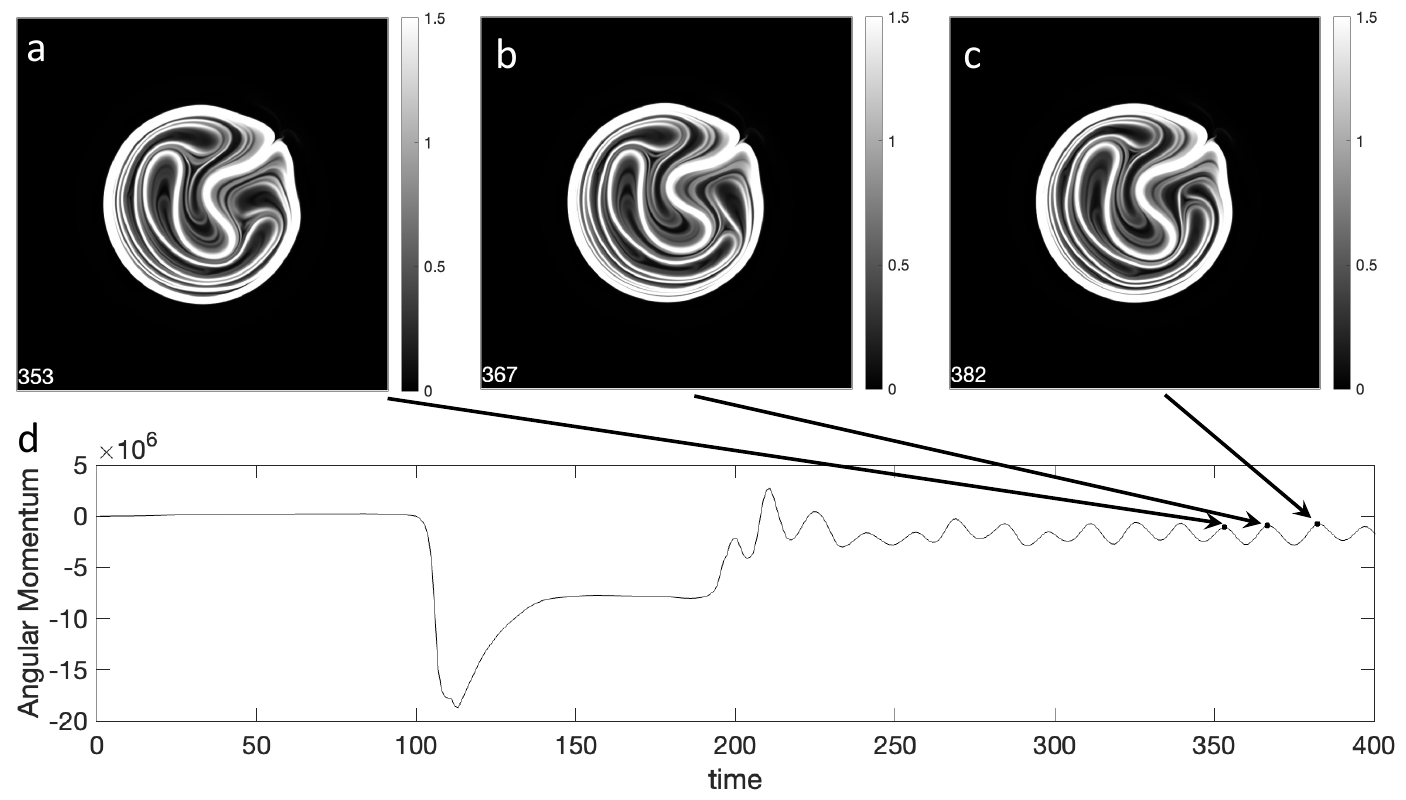}
  \caption{Simulation of the flagellar state experimentally observed in Ref.~\onlinecite{Hardouin22}.  (a) The flagellar state injecting material into the well interior.  (b) and (c) are periodic repeats of state (a).  (d) Plot of the angular momentum showing the periodic behavior of the flagellum.}
  \label{fig:Flagellar}
\end{figure}

Second, we observe the periodic ``flagellar'' state seen experimentally in Ref.~\onlinecite{Hardouin22} using the same setup as Fig.~\ref{fig:dblPer} with the exception of an $800 \times 800$ grid, $\Delta x = 0.25$, total mass $11,646$ and a smaller bulk viscosity $\xi = 18,750$.  (Fig.~\ref{fig:Flagellar} and Video S4).   Here, the material again forms a boundary ring, with the material now continuously flowing into the center of the well.  This inflow starts at a -1/2 defect at the boundary and forms a long filament, reminiscent of a flagella, that whips back and forth periodically, which is evident in Figs.~\ref{fig:Flagellar}a, b, and c and the time series of the angular momentum (Fig.~\ref{fig:Flagellar}d).

To summarize, a simple principles based model of MT active nematics produces key features of the density variations of experimental systems in both open and confined geometries.

\acknowledgments

We would like to acknowledge helpful conversations with Cristina Marchetti, Dimitrios Krommydas, Zvonimir Dogic, and Mattia Serra.  We also thank Linda Hirst for sharing experimental data.  This work was financially supported by the US Department of Energy under grant DE-SC0025803, by the US National Science Foundation through the Center of Research Excellence in Science and Technology: Center for Cellular and Biomolecular Machines at the University of California Merced (HRD-1547848 and HRD-2112675) and through NSF grant DMR-2225543, and finally by the University of California Office of the President under grant M25PL8991 (the UC Active Matter Hub).  We used the Claude LLM by Anthropic for coding assistance and implementing the numerics, as well as background research.

\onecolumngrid
\pagebreak

{\centering \LARGE Modeling density variations in two-dimensional microtubule-based active nematics: Supplemental Material}

\section{Derivation of Eq.~(\ref{r7}): the flow coupling for $\mathsf{P}$}

We follow the structure of the derivation in Ref.~\onlinecite{Mitchell25}, which assumed incompressibility, but now allow $\mathbf{u}$ to be compressible.  First, if we relax the constraint that $|\mathbf{n}| = 1$, then a tangent vector $\mathbf{n}$ evolves in a flow $\mathbf{u}$ according to 
\begin{equation}
    \frac{D}{Dt}\mathbf{n} = \mathbf{n} \cdot \nabla \mathbf{u}.
\end{equation}
Defining $\mathsf{T} = \mathbf{n} \otimes \mathbf{n}$, the above implies
\begin{align}
\frac{D}{Dt} \mathsf{T} &= \left( \frac{D}{Dt}\mathbf{n}\right)\otimes \mathbf{n} + \text{Tr.} = (\mathbf{n} \cdot \nabla \mathbf{u} )\otimes \mathbf{n} + \text{Tr.} \\
& = (\nabla \mathbf{u} )^T\mathbf{n}\otimes \mathbf{n} + \text{Tr.}
=(\mathsf{E} - \mathsf{\Omega})\mathsf{T} + \text{Tr.} \\
& = \mathsf{ET} + \mathsf{TE} +[\mathsf{T},\mathsf{\Omega}], \label{r24}
\end{align}
where ``Tr.'' denotes the transpose of what precedes it.  For a 2D flow, the following holds
\begin{equation}
    \chi \equiv \mathsf{E'T} + \mathsf{TE'} - \text{Tr}(\mathsf{TE'})\mathsf{I} - \text{Tr}(\mathsf{T})\mathsf{E}'= 0,
    \label{r23}
\end{equation}
where $\mathsf{E}'$ is the deviatoric part of $\mathsf{E}$.  To prove this equality, we first note that $\chi$ as defined above is a traceless symmetric $2\times 2$ matrix.  Such matrices are spanned by $\mathsf{P}$ and $\mathsf{R}$.  It is straightforward to show that $\text{Tr}(\chi \mathsf{P}) = \text{Tr}(\chi \mathsf{R}) = 0$, thereby proving Eq.~(\ref{r23}).  Applying Eq.~(\ref{r23}) to Eq.~(\ref{r24}) yields,
\begin{equation}
    \frac{D}{Dt} \mathsf{T} = \text{Tr}(\mathsf{T})\mathsf{E}' +[\mathsf{T},\mathsf{\Omega}] + \text{Tr}(\mathsf{TE'}) \mathsf{I} + \text{Tr}(\mathsf{E}) \mathsf{T}.
\end{equation}
Now, imposing the constraint that $|\mathbf{n}| = 1$, i.e. $\text{Tr}(\mathsf{T}) = 1$, the above becomes
\begin{equation}
    \frac{D}{Dt} \mathsf{T} = \mathsf{E}' +[\mathsf{T},\mathsf{\Omega}] + \text{Tr}(\mathsf{TE'}) (\mathsf{I} - 2 \mathsf{T}).
\end{equation}
Applying this to $\mathsf{P} = \mathsf{T} - \mathsf{I}/2$ yields the desired result
\begin{equation}
    \frac{D}{Dt} \mathsf{P} = \mathsf{E}' +[\mathsf{P},\mathsf{\Omega}] - 2 \text{Tr}(\mathsf{PE'})\mathsf{P}.
\end{equation}

\section{Derivation of the elastic stress tensor Eq.~(\ref{r18})}

The rate of change of the free energy equals the integral of the power density, expressed as velocity times force density,
\begin{equation}
    \frac{dF}{dt} = - \int u_i \nabla_j \Pi^E_{ij} \;da 
    = \int (\nabla_j u_i) \Pi^E_{ij} \; da.
    \label{r16}
\end{equation}
Using the chain rule for functional derivatives, we can also write
\begin{align}
    \frac{dF}{dt} &= \int  \left(\frac{\delta F}{\delta P_{ij}} \frac{\partial P_{ij}}{\partial t} + \frac{\delta F}{\delta \rho} \frac{\partial \rho}{\partial t}\right)\; da \nonumber \\
    & = \int \frac{\delta F}{\delta P_{ij}}\left( \frac{D}{Dt} - \mathbf{u} \cdot \nabla \right)  P_{ij}\; da + \nonumber \\ 
    & \; \;\;\;\int \frac{\delta F}{\delta \rho}\left( \frac{D}{Dt} - \mathbf{u} \cdot \nabla \right)  \rho \; da.
    \label{r13}
\end{align}
The first term in the integrand above yields
\begin{align}
    \frac{\delta F}{\delta P_{ij}} &\frac{D}{Dt}  P_{ij} = 
     -h_{ij}\frac{D}{Dt}  P_{ij} \nonumber \\
     &=
      -h_{ij}
      (\mathsf{E}' +  [\mathsf{P}, \Omega ] -
2 \text{Tr} (\mathsf{P} \mathsf{E}' ) \mathsf{P})_{ij} \nonumber \\
     &= (\nabla_j u_i) 
     (-\mathsf{h} +  [\mathsf{P}, \mathsf{h} ] +
2 \text{Tr} (\mathsf{P} \mathsf{h} ) \mathsf{P})_{ij},
\label{r14}
\end{align}
where the first equality follows from the definition of the $\mathsf{h}$-tensor Eq.~(\ref{r5}), the second equality from the nematic transport equation (\ref{r7}), and the third equality from rearranging terms.  The second term in Eq.~(\ref{r13}) yields
\begin{align}
    &-\int \frac{\delta F}{\delta P_{ij}}  \mathbf{u} \cdot \nabla   P_{ij}\; da \nonumber \\
    &= 
    -\int \left( \frac{\partial f}{\partial P_{ij}} - \nabla_k \frac{\partial f}{\partial \nabla_k P_{ij}} \right)  \mathbf{u} \cdot \nabla   P_{ij}\; da \nonumber \\
    &= - \int \left[ \frac{\partial f}{\partial P_{ij}}  \mathbf{u} \cdot \nabla   P_{ij}+ \frac{\partial f}{\partial \nabla_k P_{ij}} \mathbf{u} \cdot\nabla(\nabla_k P_{ij})\right] \; da \nonumber \\
    & - \int (\nabla_j u_i) \nabla_i P_{ij} \frac{\partial f}{\partial \nabla_j P_{ij}}   \; da 
    \label{r15}
\end{align}
where the first equality derives from the formula for the functional derivative and the second equality from a combination of the chain rule and integration by parts. 

The third term in the integrand of Eq.~(\ref{r13}) yields
\begin{equation}
    \frac{\delta F}{\delta \rho} \frac{D}{Dt} \rho = 
    -\frac{\delta F}{\delta \rho}\rho\nabla \cdot \mathbf{u} =
    (\nabla_j u_i)\rho g \delta_{ij}
\end{equation}

The fourth term in Eq.~(\ref{r13}) yields
\begin{align}
    &-\int \frac{\delta F}{\delta \rho}  \mathbf{u} \cdot \nabla \rho\; da \nonumber \\
    &= 
    -\int \left( \frac{\partial f}{\partial \rho} - \nabla_k \frac{\partial f}{\partial \nabla_k \rho} \right)  \mathbf{u} \cdot \nabla \rho\; da \nonumber \\
    &= - \int \left[ \frac{\partial f}{\partial \rho}  \mathbf{u} \cdot \nabla  \rho + \frac{\partial f}{\partial \nabla_k \rho} \mathbf{u} \cdot\nabla(\nabla_k \rho)\right] \; da \nonumber \\
    & - \int (\nabla_j u_i) \nabla_i \rho \frac{\partial f}{\partial \nabla_j \rho}   \; da 
\end{align}

Plugging Eqs.~(\ref{r14}) and (\ref{r15}) into Eq.~(\ref{r13}) yields
\begin{align}
    \frac{dF}{dt} &= \int (\nabla_j u_i) \biggl[-h_{ij} +  [\mathsf{P}, \mathsf{h} ]_{ij} +
2 \text{Tr} (\mathsf{P} \mathsf{h} ) P_{ij}\nonumber \\
    & -\nabla_i P_{k \ell} \frac{\partial f}{\partial \nabla_j P_{k \ell}} -\nabla_i \rho \frac{\partial f}{\partial \nabla_j \rho} 
  +(f + \rho g) \delta_{ij} \biggr] \; da
\label{r17}
\end{align}
Comparing Eq.~(\ref{r16}) to Eq.~(\ref{r17}) yields Eq.~(\ref{r18}) in the main text.

Note that the pressure $p = -\text{Tr} (\Pi^E)/2$ is
\begin{equation}
p = -f - \rho g  + \frac{1}{2} \nabla_k P_{ij} \frac{\partial f}{\partial \nabla_k P_{ij}} + \frac{1}{2}\nabla_k \rho \frac{\partial f}{\partial \nabla_k \rho}.
\end{equation}

\section{Numerical implementation notes}

The main simulation code is written in Python using the pytorch package to take advantage of the GPU speed up.
We numerically evolve and store $\rho$ and $\mathsf{P}$ separately, rather than combining them into $\mathsf{Q}$.  This makes it easier to impose mass conservation on $\rho$.  To that end, we advect $\rho$ using a 5th order WENO scheme with flux limiting to ensure that $\rho$ remains positive.  We also use Kahan compensation to minimize mass loss from round-off error.

We evolve $\mathsf{P}$ via the equation $\partial \mathsf{P} / \partial t= (\partial \theta / \partial t) \mathsf{R}$, where $\theta$ is the angle of the director field $\mathbf{n}$.  We compute $\partial \theta/\partial t$ by taking the trace of Eq.~(\ref{r11}) multiplied by $\mathsf{R}$.  Special attention needs to be paid to the advection part of Eq.~(\ref{r11}), which is achieved using an upwind finite difference technique.  Once $\partial \theta/ \partial t$ is computed, $\mathsf{P}$ is evolved using a rotation matrix to preserve the norm of $\mathsf{P}$.

To evolve $\mathsf{u}$, we could solve the Stokes equation (\ref{r12}) directly.  However, we opt to evolve $\mathbf{u}$ using the full Navier-Stokes form
\begin{equation}
   \rho_0 \frac{D}{Dt}u_i = \rho_0\left ( \frac{\partial}{\partial t}u_i + \mathbf{u} \cdot \nabla u_i \right )  = \nabla_j\Pi_{ij},
    \label{r25}
\end{equation}
where $\rho_0$ is the total density advected, equal to the sum of the density of the microtubules $\rho$ and the water entrained with them.  We assume for simplicity that this total is constant (i.e. $\rho_0 = 1$).
We use an explicit scheme to incorporate all of the terms except the shear and bulk viscosity for which we use an implicit FFT scheme.  Again, special care must be taken with the advection term $\mathbf{u} \cdot \nabla u_i$, for which we again use upwinding.

When the potential well is utilized, there is the question of how to handle the fields in the ``unphysical'' region outside the well.  The simplest approach is to treat this region as the rest of the domain, relying on the potential function to keep the mass inside the well, and letting the fields $\mathbf{u}$ and $\mathsf{P}$ evolve as they will outside the well.  This approach generally works, but defects tend to proliferate outside the well.  We thus implement two numerical techniques to suppress dynamics outside the well.  First, we introduce a velocity-dependent friction term of the form $-\mu_f \mathbf{u}$ added to $\partial \mathbf{u}/ \partial t$ in Eq.~(\ref{r25}).  This term is only applied outside the well, gated by the normalized potential function squared, $(V_{\text{well}}(r)/V_0)^2$.  Second, we introduce a smoothing to $\mathsf{P}$ outside the well to reduce defect proliferation.  This is achieved by adding a diffusion term $\mu_d \nabla^2 \mathsf{P}$ to $\partial \mathsf{P}/ \partial t$.  This term is again gated by $(V_{\text{well}}(r)/V_0)^2$.  Furthermore, after the diffusion is applied to $\mathsf{P}$, it is renormalized so that $\text{Tr}(\mathsf{P}^2) = 1/2$.

\section{Varying Parameters}
Two key parameters are introduced into this new model: the MT density $\rho$ and the bulk viscosity $\xi$. Recall the free energy introduced in Eq.~\ref{r10} and the active stress $\Pi^A = - \zeta \rho^a (\mathsf P + \mathsf I/2)$. By absorbing the prefactors in the free energy into a constant $K' = K\rho^b$, and similarly with the activity $\zeta ' = \zeta \rho^a$, we can derive an expression for the active length $\ell_a$, generally defined to be $\ell_a = \sqrt{K/\zeta}$. The active length for this model is $\ell_a' = \sqrt{K'/\zeta'} = \rho^{(b-a)/2} \sqrt{K/\zeta}$. Thus, scaling density by $c$ scales the active length by $c^{(b-a)/2}$. 

To observe the effects of varying bulk viscosity $\xi$, we performed three simulations. The simulations were done with the parameters in Table~\ref{tab:default-values}, with varying values of $\xi$. Images of the three simulations can be found in Fig~\ref{fig:bulk-visc-comparison}. Decreasing the bulk viscosity to $\xi = 2,560$ increases the compressibility of the bulk material, causing the MT bundles to coalesce into high density ribbon-like bands, leaving large low-density voids. Increasing the bulk viscosity to $\xi = 75,000$ decreases the compressibility, which suppresses the formation of those high density bands and instead produces a comparatively more spatially uniform density field. Note that the color bar in each panel in Fig.~\ref{fig:bulk-visc-comparison} is bounded by the global extrema of each simulation, so changing $\xi$ also affects the contrast between the minimum and maximum density.

\section{Summary of Simulation Parameters}

The grid spacing is taken into account in the simulations by appropriately scaling the physical parameters with length dependence: $K' = K/\Delta x^2$, $\eta' = \eta/\Delta x$, $\xi' = \xi/\Delta x$, $\Gamma' = \Gamma \Delta x$,
$R_{\text{Well}}' = R_{\text{Well}}/\Delta x$, $W_{\text{Well}}' = W_{\text{Well}}/\Delta x$.  Note that we do not scale the switch function values $\kappa^2_c$ and $\kappa^2_w$.  With these rescaled values, the simulation is then run using $\Delta x = 1$.

\begin{table}
  \centering
  
  \begin{tabular}{|cc|c|}
    \hline
           &  & Default \\
    \hline
    basics   & grid & $200 \times 200$  \\
       & $\Delta x$ &  1 \\
       & $m$ &  40,000 \\
\hline
    elasticity  & $K$ &  90112 \\
       & $b$ & 2  \\
\hline
    activity  & $\zeta$ &  4096 \\
       & $a$ &  2.4 \\
\hline
    viscosity  & $\eta$ & 2560 \\
       & $\xi$ &  18,750 \\
       & $\Gamma$ & 1 \\
\hline
    switch & $\rho_c$ & 0.3 \\
       & $\rho_w$ &  0.03 \\
       & $\kappa^2_\ell$ &  0.025\\
       & $\kappa^2_u$ &  0.4\\
       & $\kappa^2_w$ &  0.0025\\
\hline
    potential & $R_{\text{well}}$ & 60 \\
       & $W_{\text{well}}$ &  2  \\
       & $V_0$ &  100,000\\
\hline
    external & $\mu_f$  & 1000 \\
       & $\mu_d$ & 0.2  \\
    \hline
  \end{tabular}
    \caption{Default values for the model parameters}
    \label{tab:default-values}
\end{table}

\begin{table*}
  \centering
  \label{tab:mytable}
  \begin{tabular}{|cc|c|c|c|c|c|c|}
    \hline
           &  & Initial run & Low bulk visc. & Medium bulk visc. & High bulk visc. & Doubly periodic run & Flagellar run \\
    \hline
       & Output & Fig.~\ref{fig:Exp_Simulations}c, Video S1 and S2 & Fig.~\ref{fig:bulk-visc-comparison}a & Fig.~\ref{fig:bulk-visc-comparison}b & Fig.~\ref{fig:bulk-visc-comparison}c & Fig.~\ref{fig:dblPer}, Video S3 & Fig.~\ref{fig:Flagellar}, Video S4\\
       & grid & $800 \times 800$ & $200 \times 200$& $200 \times 200$& $200 \times 200$ & $200 \times 200$ & $800 \times 800$\\
       & $\Delta x$ &  0.25 & 1& 1& 1& 1 & 0.25\\
       & $m$ &  40,000 & 40,000& 40,000& 40,000& 4,709  &  11,646\\
       & $\xi$ &  75,000 & 2,560 & 25,600 & 75,000 & 75,000 & 18,750\\
\hline
  \end{tabular}
    \caption{Model parameters for specific runs}
\end{table*}

\section{Captions to Supplemental Videos}

S1: Video of simulation in Fig.~\ref{fig:Exp_Simulations}c.

S2: Video of simulation in Fig.~\ref{fig:Exp_Simulations}c with defects added.

S3: Video of doubly periodic state in Fig.~\ref{fig:dblPer}.

S4: Video of flagellar state in Fig.~\ref{fig:Flagellar}.

\end{document}